\documentclass[pop,superscriptaddress,amsmath,twocolumn]{revtex4}  

\usepackage{graphicx}
\usepackage{gensymb}
\usepackage{dcolumn}
\usepackage{bm}
\usepackage{color}
\usepackage{amsfonts, amsmath, bm, amssymb,graphicx}

\begin{document}

\newcommand{\beq}{\begin{equation}}
\newcommand{\enq}{\end{equation}}
\newcommand{\half}{\hbox{$1\over2$}}
\newcommand{\bmu}{{\bf\mu}}
\newcommand{\upi}{{\pi}}

\newcommand\Real{\mbox{Re}} 
\newcommand\Imag{\mbox{Im}} 
\newcommand\Rey{\mbox{\textit{Re}}}  
\newcommand\Pran{\mbox{\textit{Pr}}} 
\newcommand\Pen{\mbox{\textit{Pe}}}  
\newcommand\Ai{\mbox{Ai}}            
\newcommand\Bi{\mbox{Bi}}            
\newcommand\E{\mbox{e}}
\newcommand\I{\mbox{i}}
\newcommand\D{\mbox{d}}

\newcommand{\C}{\mathbb C}
\newcommand{\e}{\eqref}
\newcommand{\Lap}[1]{\Delta^{\!\!(#1)}}
\newcommand{\lb}{\left(}
\newcommand{\rb}{\right)}
\newcommand{\lsb}{\left[}
\newcommand{\rsb}{\right]}
\newcommand{\Rin}{R_{i}}
\newcommand{\Xout}{X_{o}}
\newcommand{\Xin}{X_{i}}
\newcommand{\Rout}{R_{o}}
\newcommand{\Lin}[1]{\left\|#1\right\|_{i}}
\newcommand{\Lout}[1]{\left\|#1\right\|_{o}}
\newcommand \mcE{{\mathcal E}}
\newcommand \bfr{{\bf r}}
\newcommand \bfk{{\bf k}}
\newcommand \bfq{{\bf q}}
\newcommand \bfU{{\bf U}}
\newcommand \bfV{{\bf V}}

\def\res{\mathop{\mathrm{Res}}\limits}
\newcommand{\sgn}{\operatorname{sgn}}
\newcommand{\R}{\mathbb R}
\newcommand{\p}{\partial}
\newcommand\Fr{\mbox{\textit{Fr}}}
\newcommand\Ham{\mbox{\textit{H}}}
\newcommand\Arg{\mbox{\textrm{Arg\,}}}
\newcommand\bigO{\mbox{\mathcal{O}}}


\title{The instability of near-extreme Stokes waves}

\author{Bernard Deconinck}
\affiliation{Department of Applied Mathematics, University of Washington,
Seattle, WA 98195-3925, USA}
\author{Sergey A. Dyachenko}
\affiliation{Department of Mathematics, SUNY Buffalo, Buffalo NY,
14260-2900, USA}
\author{Pavel M. Lushnikov}
\affiliation{Department of Mathematics \& Statistics, University of New
Mexico, Albuquerque, NM 87131, USA}
\author{Anastassiya Semenova}
\affiliation{Department of Applied Mathematics, University of Washington,
Seattle, WA 98195-3925, USA}

\date{\today}

\begin{abstract}
    We study the stability of Stokes waves on a free surface of an ideal fluid of infinite depth.
    For small steepness the modulational instability dominates the dynamics, but its growth rate is vastly surpassed for steeper waves by an instability due to disturbances localized at the wave crest, explaining why long propagating ocean swell consists of small-amplitude waves. The dominant localized disturbances are either co-periodic with the Stokes wave, or have twice its period.
    The nonlinear stage of instability for steep wave evolution reveals the formation of a plunging breaker. 
\end{abstract}

\maketitle



%
Ocean swell in the wake of a storm or driven by the force of strong wind is seen as a collection of almost periodic unidirectional waves from a beach or an ocean liner.
The waves that do not vary in the direction transverse to propagation are known as Stokes waves. 
Such waves were discovered by Stokes~\cite{stokes1847theory} who  conjectured~\cite{Stokes1880_314} that the wave of greatest height, referred to as the extreme Stokes wave, has a crest angle of $120$\degree. 
That conjecture was proven
more than 100 years later, see \cite{AmickFraenkelTolandActaMath1982, PlotnikovDinamikaSploshnSredy1982}. 
The study of Stokes waves has been the subject of numerous works, see e.g.~\cite{michell1893,Nekrasov1921,MalcolmGrantJFM1973LimitingStokes,SchwartzJFM1974,Longuet-HigginsFoxJFM1977,Longuet-HigginsFoxJFM1978,toland1978existence,PlotnikovDinamikaSploshnSredy1982,AmickFraenkelTolandActaMath1982,Williams1981,WilliamsBook1985,TanveerProcRoySoc1991,gandzha2002steep,Longuet-HigginsWaveMotion2008,DyachenkoLushnikovKorotkevichJETPLett2014,DyachenkoLushnikovKorotkevichPartIStudApplMath2016,LushnikovStokesParIIJFM2016,LushnikovDyachenkoSilantyevProcRoySocA2017}.
Stokes waves are a one-parameter family of solutions often characterized by the steepness $s=H/L,$ where $H$ is the crest-to-trough height and $L$ is the wave length. 
The steepness of the extreme Stokes wave is reached at $s_{lim}=0.14106348398\ldots$~\cite{LushnikovDyachenkoSilantyevProcRoySocA2017} (see also~\cite{dhs2022almost} for a more accurate value).
However, the curvature radius at the crest of the Stokes wave vanishes as the extreme wave is approached. In the ocean, once the capillary length scale $\simeq 1.7$cm (for ocean water) is reached, the surface tension effects cannot be disregarded.
This limits the physical relevance of Stokes waves in the immediate vicinity of the extreme wave.
Nevertheless, the wavelength of gravity waves in ocean swell often ranges from a dozen centimeters to a few kilometers, see e.g.~\cite{YoungBookOceanWaves1999,liu2003effects}. 
Thus, without oversimplification, a wide range of Stokes waves is relevant for modeling oceanic swell.
The Stokes waves found in~\cite{DyachenkoLushnikovKorotkevichPartIStudApplMath2016,LushnikovDyachenkoSilantyevProcRoySocA2017} were used to determine the radius of curvature at the crest. 
Surface tension effects are small for $s \lesssim s_{50} = 0.14100$ and $s\lesssim s_{100} = 0.141034$ for Stokes waves 
with wave length $50$ and $100$ meters respectively. 
In what follows we provide stability results for $s\le  0.1408627$ that are well below these applicability estimates. 
The nonlinear stage of instability, as observed in experiments or computation, reveals that wave breaking leads to the rapid vanishing of the curvature radius to the point that surface tension becomes non-negligible, and whitecaps form through ejection of capillary waves from spilling/plunging breakers, see~\cite{longuet1963generation,duncan2001spilling,dyachenko2016whitecapping}.

\begin{figure}
\includegraphics[width=0.495\textwidth]{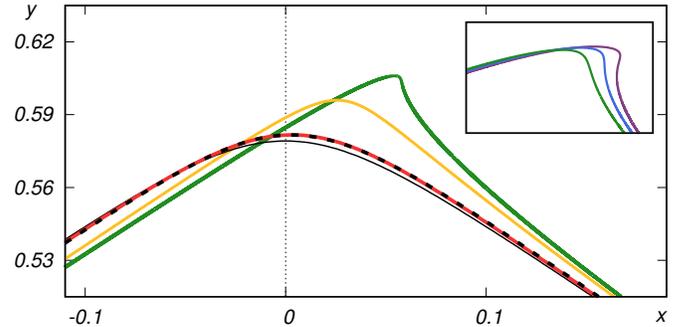}
\vspace*{-0.25in}
\caption{Initial stage of instability development at the crest of a Stokes wave (solid black) from the nonlinear dynamics (red line) compared to a perturbation of the Stokes wave by the eigenfunction corresponding to a localized instability (black dashed line). The later stage of the dynamics reveals the formation of a breaker (orange \& green) and overturning (inset).
}
\vspace*{-0.25in}
\label{fig:breaking}
\end{figure}

Since \cite{KharifRamamonjiarisoaPhysFluids1988} indicates that transverse perturbations of steep waves have smaller growth rates than do longitudinal ones, 
we examine the growth rates of longitudinal instabilities of Stokes waves in water of infinite depth, ignoring the effects of surface tension, vorticity, wind and dissipation. 
We consider {\em all} quasiperiodic perturbations, {\em i.e.}, we include {\it superharmonic}  (with period $L$) and {\it subharmonic} (with larger period, an integer multiple of $L$) perturbations, using the terminology of~\cite{Longuet-HigginsFoxProcRoySoc1978superharmonics, Longuet-HigginsFoxProcRoySoc1978subharmonics}. 
The subharmonic instability of small-amplitude Stokes waves (in water waves and other fields) has been studied since 1965, see Refs.~\cite{benjamin1967disintegration, bridgesmielke, Lighthill1965,whitham1967non,Zakharov1968,ZakharovOstrovskyPhysD2009} and is now well understood \cite{berti2021full, creedon2022ahigh, nguyenstrauss}.
 \begin{figure*}
    \centering
    \includegraphics[width=0.364\textwidth]{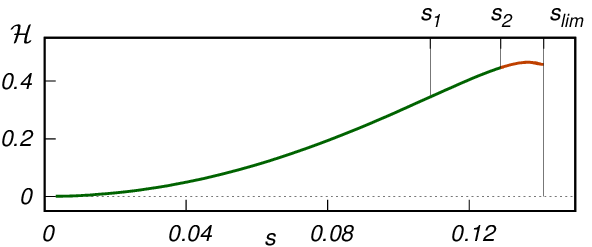}
    \includegraphics[width=0.621\textwidth]{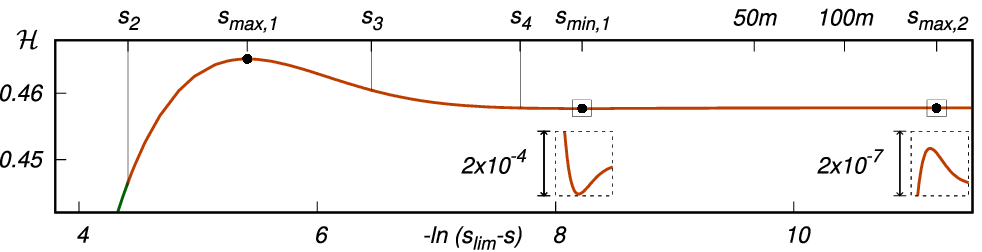}
    \vspace*{-0.1in}
    \caption{(Left panel) Hamiltonian $\cal H$ given by~\eqref{HamConf} as a function of steepness $s$. Green denotes the region of dominance of the BF instability, red corresponds to dominance of the localized branch.
    (Right panel) $\cal H$ as a function of $-\ln\left( s_{lim} - s\right)$ oscillates near the extreme wave. The points marked $s_{max,1},s_{min,1},s_{max,2}$ are the first three extrema of $\cal H$. For details about $s_1, s_2, s_3$ and $s_4$ we refer the reader to the main body of the text. 
    }
    \label{fig:ham} 
    \vspace*{-0.2in}
\end{figure*}
This instability is referred to as the Benjamin-Feir (BF) or modulational instability.
The growth rate of the BF instability increases with $s$, passes through its maximum and decays to zero.
The so-called {\em localized} instability branch 
appears for steep waves before the BF instability ceases.
The eigenfunctions associated with this branch change rapidly in the vicinity of the wave-crest. 
Increasing $s$ gives rise to additional localized instability branches.
These branches correspond to both the sub-
\cite{Longuet-HigginsFoxProcRoySoc1978subharmonics,McLeanMaMartinSaffmanYuenPRL1981,KharifRamamonjiarisoaPhysFluids1988}
and the superharmonic disturbances (the latter form at the extrema of the Hamiltonian, {see Fig.~\ref{fig:ham}})
\cite{Longuet-HigginsFoxProcRoySoc1978superharmonics,tanaka1983stability,LonguetHigginsTanakaJFM1997,KorotkevichLushnikovSemenovaDyachenkoStudApplMath2022accepted}.
See also \cite{KharifRamamonjiarisoaPhysFluids1988,KorotkevichLushnikovSemenovaDyachenkoStudApplMath2022accepted} for earlier numerical studies of different branches.

It has been unclear as to which of the instabilities dominates the dynamics of almost-extreme Stokes waves. 
The present work elucidates this long-standing question and establishes the localized branch as the dominant one for steep waves.
Depending on the value of $s$, either a perturbation co-periodic with the Stokes wave, or the one having twice its period is the most unstable.
The dominance of the latter perturbations interchanges repeatedly as the extreme wave is approached.
Growth rates observed for near-extreme waves achieve large values, resulting in instabilities that would swiftly disintegrate wave trains of large-amplitude Stokes waves. 
This is crucial for understanding why long-lived oceanic swell is never observed in the ocean.
%
%
%
%


\begin{figure*}
    \centering
    \includegraphics[width=0.395\textwidth]{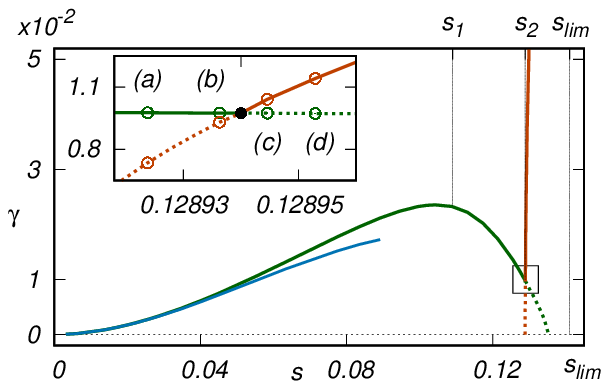}
    \includegraphics[width=0.597\textwidth]{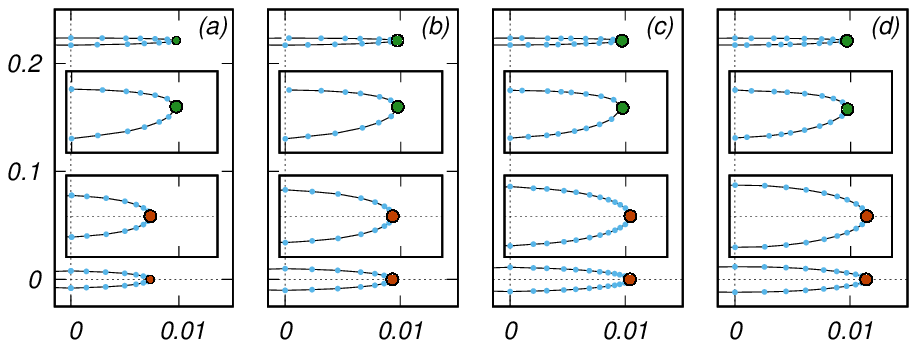}
    \vspace*{-0.25in}
    \caption{(Left panel) The maximal instability growth rate $\gamma$ as a function of steepness $s$. BF (green) is the dominant instability for small steepness, and the localized branch (red) dominates for large steepness. The blue curve is the theoretical curve from~\cite{creedon2022ahigh}.
    The BF figure eight detaches at steepness $s_1$ and begins to shrink. At $s=s_2$, the dominant instability transitions to the localized branch with real eigenvalue. The inset shows how the 
    transition to the localized branch occurs. The spectral plane 
    for waves $(a)$-$(d)$ is shown on the right.
    (Right panel) Spectrum of the problem (real part vs. imaginary part) in the transition region $(a)$-$(d)$ from the inset of the left panel. Red circles correspond to the subharmonic instability ($\mu = 0.5$) of the localized branch and green circles are the eigenvalues with largest real part of the BF remnant.
    }
    \vspace*{-0.1in}
    \label{fig:growth}
\end{figure*}

\noindent {\bf Problem Formulation}. We consider potential motion of an ideal fluid in a 2D domain, ${\bf r} = (x,y)^T\in\mathcal{D}(t) \subset \mathbb{R}^2$.
The fluid velocity ${\bf v}$ is the gradient of the potential $\varphi({\bf r},t)$: ${\bf v}= \nabla \varphi$ with $\nabla\equiv(\p_x,\p_y)$ which satisfies $\nabla^2 \varphi = 0$.  
The fluid domain is bounded by a $2\pi$-periodic free surface and extends downward to infinity. 
The Hamiltonian of the Euler equations with a free surface is recast in terms of surface variables~\cite{Zakharov1968}, and the conformal mapping approach from \cite{TanveerProcRoySoc1991,DyachenkoZakharovKuznetsovPlasmPhysRep1996} is used. 
Here $w=u+iv\in \mathbb{C}^-$ parameterizes the fluid domain through the conformal mapping $z(w,t) = x(w,t) + i y(w,t) \in \mathcal{D}(t)$.  

The Hamiltonian of the fluid flow is the sum of the kinetic and  potential energy due to gravity. 
The Eulerian dynamics with a free surface is obtained from the least action principle~\cite{DyachenkoEtAl1996}. 
The Hamiltonian in conformal 
variables is
\begin{align}
    \mathcal{H} = -\frac{1}{2}\int\limits_{-\pi}^\pi \psi \hat H \psi_u \,du + \frac{g}{2} \int\limits_{-\pi}^\pi y^2 x_u \,du, \label{HamConf}
\end{align}
where $g$ is the acceleration of gravity, and $\hat H$ is the circular Hilbert transform. The Lagrangian has the form
\begin{align}
\mathcal{L} = \int\limits_{-\pi}^\pi \psi \left(y_t x_u - y_u x_t\right)\,du - \mathcal {H}.
\end{align}
Varying the action $\int \mathcal{L}\,dt$, with respect to $x$, $y$ and $\psi$, holding the Cauchy-Riemann relations for $x$ and $y$ as a constraint, one arrives at the equations of motion~\cite{DyachenkoEtAl1996,DyachenkoLushnikovZakharovJFM2019}:
\begin{align}
\label{impbase}
\begin{array}{l}
y_t x_u - y_u x_t = -\hat H\psi_u,  \\ 
x_t \psi_u \!-\! x_u \psi_t \!-\!\hat H\left[y_t \psi_u -  y_u\psi_t \right] =
g(x_u y \!-\! \hat H \left[yy_u\right]),
\end{array}
\end{align}
where $x = u - \hat H y$.
 It has been shown~\cite{DLZ1995Fivewave} that $y$ 
and $\mathcal{P} = x_u\psi - \hat H\left[y_u\psi\right]$ are canonical conformal variables.

The Stokes wave is a solution of \eqref{impbase}, traveling at constant speed $c$, measured in units of $c_0=\sqrt{g/k}$, the speed of linear gravity waves. 
The time variable is normalized to the frequency of linear gravity waves and is measured in units $(gk)^{-1/2}$.
After rescaling the spatial and conformal coordinates we set the 
period of the Stokes wave to be $2\pi$ in $u$ (and $x$) without loss of generality.
Writing 
\eqref{impbase} in the moving frame using $y(u-ct)$ and $\psi(u-ct)$, we linearize
around the Stokes wave in canonical conformal variables $y$ and $\mathcal{P}$ as in~\cite{DyachenkoSemenova2022}.
This results in a linear  integro-differential equation with variable periodic coefficients. 
Quasi-periodic solutions are sought in the form~\cite{deconinck2006computing}%
\begin{equation}\label{deltapsideltay}
\begin{pmatrix} \delta \mathcal{P}(u,t) \\
\delta y(u,t)
\end{pmatrix}=e^{i\mu u+\lambda t}\begin{pmatrix}\delta \mathcal{P}_0(u) \\
\delta y_0(u)
\end{pmatrix}
\end{equation}
where $\delta y_0(u)$ and $\delta \mathcal{P}_0(u)$ are $2\pi$-periodic functions, $\mu\in(-1/2,1/2]$ is the Floquet exponent, and $\lambda$ is a complex eigenvalue whose real part is the growth rate $\gamma = \mbox{Re}\, \lambda$ (if positive). 
The stability spectrum of a Stokes wave is found as the union of point spectra for all values of $\mu$~\cite{deconinck2006computing, kapitulapromislow}.

\begin{figure*}
    \centering
    \includegraphics[width=0.495\textwidth]{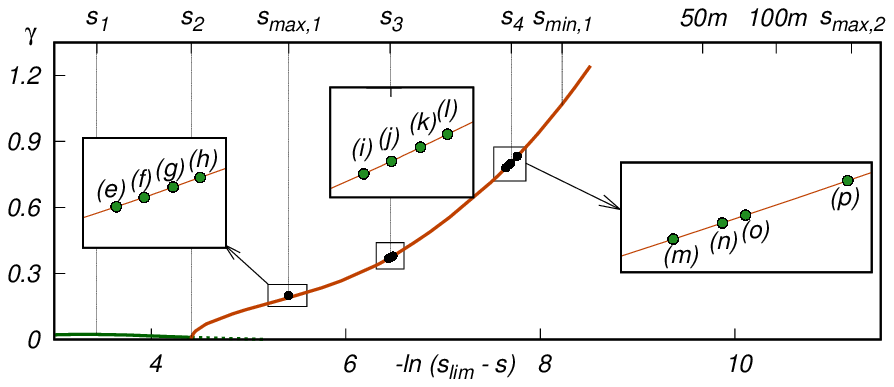}
    \includegraphics[width=0.495\textwidth]{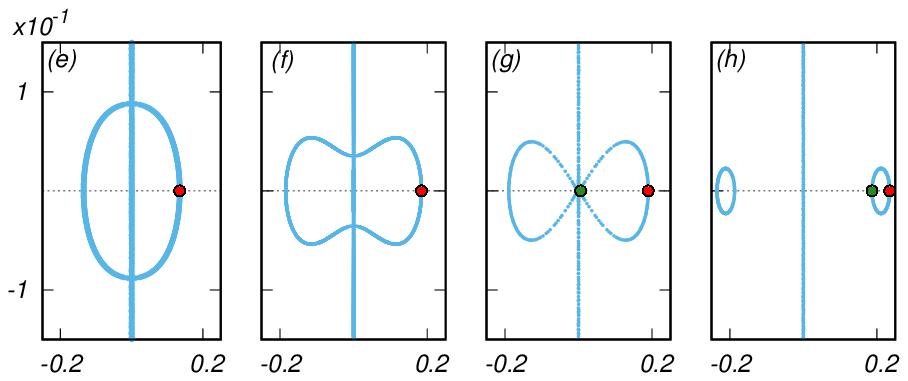}
    \vspace*{-0.25in}
    \caption{(Left panel) The maximal growth rate $\gamma$ as a function of the variable $-\ln(s_{lim}-s)$. 
    The red curve represents the localized instability with real eigenvalues for $\mu = 0$ and $\mu = 1/2$.
    At $s=s_3$, the dominant instability becomes superharmonic ($\mu=0$). At $s_4$ it switches back to subharmonic.
    (Right panel) Eigenvalues for steepness (e)-(h) in the vicinity of $s_{max,1}$: (from left to right) $s = 0.1341061714$, $s = 0.1364173038$, $s = 0.1366051511$ and $s = 0.1378879221$. As the steepness increases, superharmonic eigenvalues (green) move along the imaginary axis toward the origin and collide. As a result a pair of symmetric instability isole (h) appear and move away from the origin.
    }
    \vspace*{-0.2in}
    \label{fig:all}
\end{figure*}

The fast Fourier transform (FFT) is used to apply the linearization operator to the periodic functions appearing in the eigenvalue problem.
The resulting spectral problem in canonical conformal variables $y$ and $\mathcal{P}$~\cite{DyachenkoSemenova2022} is amenable to the shift-invert method, applied 
in $O(m N \ln N)$ flops, where $N$ is the number of Fourier 
modes and $m$ is the number of iterations to solve the linear system embedded in the shift-invert operator,
using the MINRES method~\cite{saad1992numerical}. Combined with the Krylov-Schur algorithm~\cite{lehoucq1989arpack}, the numerical solution of the eigenvalue problem is solved with up to $N = 10^{6}$ Fourier modes.

\noindent {\bf Main Results.}
The stability of Stokes waves of small to moderate steepness was studied 
using variations of the Fourier-Floquet-Hill method \cite{deconinck2006computing} by many authors, {\em e.g.}, \cite{DeconinckOliverasJFM2011, KharifRamamonjiarisoaPhysFluids1988,mclean1982instabilities}. 
Our results confirm that for $s < 0.1381465$, all Stokes waves are found to be unstable, see Fig.~\ref{fig:growth}. 
For higher steepness waves, our results are new for subharmonic perturbations ($\mu \neq 0$). They are in agreement with \cite{KorotkevichLushnikovSemenovaDyachenkoStudApplMath2022accepted} for superharmonic perturbations ($\mu = 0$). 

For steepness $0<s< s_2=0.12894$, we find that the dominant instability of Stokes waves is the modulational (BF) instability, corresponding to subharmonic perturbations, see Fig.~\ref{fig:growth}. The figure-$8$
detaches from the origin at $s_1 = 0.10423$, shortly after the maximum growth rate, $\gamma_{BF} = 0.02357$ is attained. At $s_2 = 0.12894$, the dominant instability switches (Fig.~\ref{fig:growth}) to the {\it localized} branch, described next. 

\begin{figure}
\includegraphics[width=0.485\textwidth]{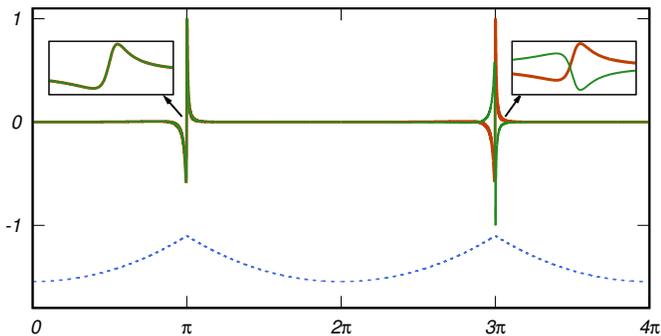}
\vspace*{-0.2in}
\caption{ 
Eigenfunctions of the localized instability branch for $\mu=0$ (red) and $\mu = 1/2$ (green) of the Stokes wave with steepness $s = 0.1406007221$ (blue). The corresponding wave is sketched in a dashed line.  Insets show zooms near peaks.
} 
\label{fig:eigf0}
\vspace*{-0.2in}
\end{figure}

The localized instability branch first appears at the steepness $s_{loc,1} = 0.12890308$ from a 
collision of eigenvalues with $\mu = 1/2$ at the origin, resulting in an oval of eigenvalues, see Fig.~\ref{fig:all}e. As $s$ increases, the oval is pinched vertically (Fig.~\ref{fig:all}f) as more eigenvalues from the imaginary axis flood onto it. This results in formation of a figure ``infinity'' at $s=s_{max,1}$ (Fig.~\ref{fig:all}g), for the Stokes wave corresponding to the first maximum of the Hamiltonian (Fig.~\ref{fig:ham}). With further increase of $s$, the figure infinity is pinched at the origin, its two isole drifting away along the real axis, as their diameter decreases (Fig.~\ref{fig:all}h). We refer to the localized instability branch as such because its eigenvalues for $\mu\in(-\frac{1}{2},\frac{1}{2}]$ have eigenfunctions that are localized near the wave crest as the extreme wave is approached, see Fig.~\ref{fig:eigf0}.  
After detachment ($s > s_{max,1}$) the entire range of the Floquet parameter $\mu \in(-\frac{1}{2}, \frac{1}{2}]$ covers the localized instability branch, in contrast to the high-frequency instabilities discussed in \cite{DeconinckOliverasJFM2011, cdo}.

\begin{figure}
\centering
\includegraphics[width=0.49\textwidth]{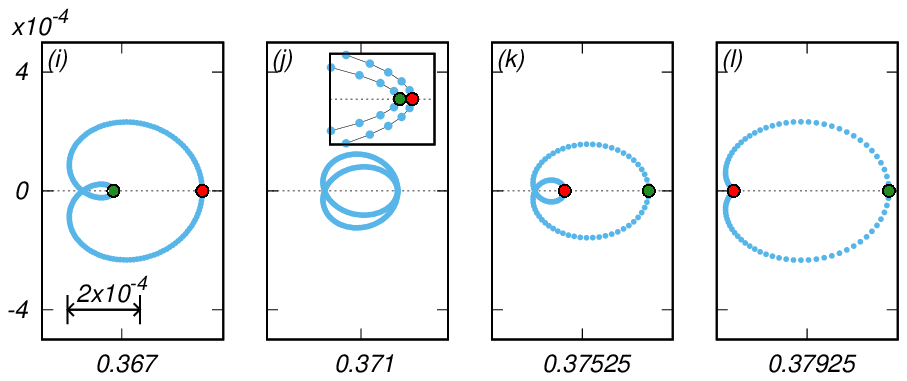}
\includegraphics[width=0.49\textwidth]{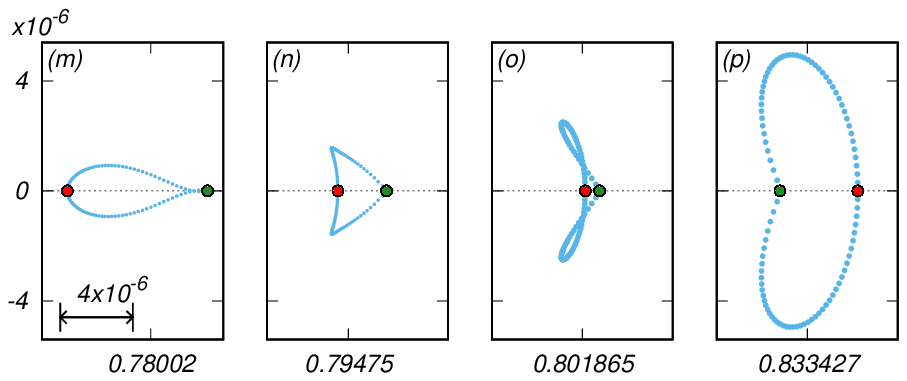}
\vspace*{-0.25in}
\caption{Spectrum in transition regions $(i)$-$(l)$ and $(m)$-$(p)$ from the
left panel of Fig.~\ref{fig:all}: around $s_3$ (top panel, from left to right) $s = 0.13946478$,
$0.13948945$, $0.13951484$ and $0.13953804$; and around $s_4$ (bottom panel,
from left to right) $s = 0.1405850778$, $0.1406007221$, 
$0.1406080087$ and $0.1406384552$. Red circles correspond to $\mu = 0.5$ (subharmonic) and green circles to $\mu=0$ (superharmonic). This demonstrates the switch between the two types of instabilities. At $s_5 = 0.14077 < s_{min,1}$, the interchange of dominance between $\mu =0$ and $\mu = 1/2$ occurs again.} 
\vspace*{-0.1in}
\label{fig:bubble1}
\end{figure}

On the unstable localized branch isola, the subharmonic mode with $\mu=1/2$ has the dominant growth rate for $s_{max,1} < s < s_3 = 0.139492$. Through a sequence of topological changes to the isola (Fig.~\ref{fig:bubble1}i-l), the superharmonic mode ($\mu = 0$) becomes dominant for $s_3 < s < s_4 = 0.140613$. For Stokes waves with larger steepness, these switches repeat. One such additional switch is illustrated in Fig.~\ref{fig:bubble1}m-p. We conjecture that the dominance of unstable mode switches between $\mu=0$ and $\mu=1/2$ an infinite number of times as the extreme wave is approached.

As $s$ increases, a secondary modulational instability branch appears and vanishes, repeating the stages of the BF instability. 
Similarly, a secondary localized branch appears through the collision of a pair of $\mu=1/2$ eigenvalues at the origin, repeating the stages seen in Fig.~\ref{fig:bubble1}i-l. The figure infinity detaches from the origin at the next extremum of the Hamiltonian at $s = s_{min,1}$. We conjecture that 
new modulational and new localized instability branches appear infinitely many times 
as the extreme wave is approached. 

Eigenfunctions on the localized instability branch are shown in Fig.~\ref{fig:eigf0} for $\mu=0$ (two periods) and $\mu=1/2$ (one period). 
As $s$ increases, the eigenfunctions are increasingly localized near the crest of the wave.  
When used as a perturbation to seed the nonlinear dynamics about a Stokes wave, the localized instability leads to the steepening and overturning of the wave, see Fig.~\ref{fig:breaking}. 
Simultaneously, the dominant unstable isola drifts to infinity, as the extreme wave is approached. An infinite growth rate is not physical, indicating the breakdown of our model. Indeed, as increasingly steep
waves approach the extreme 120\degree\, Stokes wave, the effects of
surface tension, among others, play a role and have to be incorporated~\cite{banner2014linking}. 

\noindent {\bf Conclusion}\,
While the instability spectrum is quite complicated, the instability consequences for highly nonlinear Stokes waves are straightforward. Unlike for small-amplitude waves \cite{DeconinckOliverasJFM2011}, the dominant eigenvalue is real and is associated with
eigenfunctions ({\em i.e.}, perturbations) that are co-periodic
with the Stokes wave or have double its period. In either case, the instabilities that dominate the dynamics are localized near the wave's crest. 

Our numerical observations are particularly significant because they
answer long-standing questions of which instability dominates the
evolution of steep ocean waves. Further, our computations illustrate
the complexity of the spectral stability problem governing the different
instabilities of large-steepness Stokes waves. We have focused on the
dominant instabilities of steep waves, but the stability spectrum of such
waves has many sub-dominant components, which will be discussed elsewhere.

\noindent {\bf Acknowledgements} 
The authors thank T. J. Bridges and A. O. Korotkevich for helpful discussions. 
The work of P.M.L. was supported by NSF-DMS-1814619. 
P.M.L. and S.D. thank the Isaac Newton Institute for Mathematical
Sciences, Cambridge, UK for support and hospitality during the programme ``Dispersive hydrodynamics''
where work on this paper was partially undertaken. 
S.D. thanks the Department of Applied Mathematics at the U. of Washington for hospitality. A.S and S.D. acknowledge the FFTW project and its authors~\cite{frigo2005design} as well as the entire GNU Project. 
A.S. thanks the Institute for Computational and Experimental Research in Mathematics, Providence, RI, being resident during the ``Hamiltonian Methods in Dispersive and Wave Evolution Equations" program supported by NSF-DMS-1929284.
\bibliography{StokesWave,surfacewaves_PL,lushnikov,books}

\end{document}